\documentclass{aastex62}

\received{***}
\revised{***}
\submitjournal{ApJ}
\def\vec#1{\mbox{\boldmath $#1$}}
\shorttitle{An MHD Modeling of Successive Flares of 2017 September 6}
\shortauthors{Inoue, $\&$ Bamba}
\begin{document}

\title{An MHD Modeling of Successive X2.2 and X9.3 Solar Flares of 2017 September 6}

\correspondingauthor{Satoshi Inoue}
\email{si22@njit.edu}

\author[0000-0001-5121-5122]{Satoshi Inoue}
\affil{Center for Solar-Terrestrial Research, New Jersey Institute of Technology, 323 Dr M.L.K. Jr Blvd, Newark, NJ 07102-1983, USA}
\affil{Institute for Space-Earth Environmental Research, Nagoya University Furo-cho, Chikusa-ku, Nagoya, 464-8601, Japan}

\author[0000-0002-0786-7307]{Yumi Bamba}
\affil{Institute for Advanced Research, Nagoya University, Furo-cho Chikusa-ku Nagoya 464-8601, Japan}
\affil{Institute for Space-Earth Environmental Research, Nagoya University Furo-cho, Chikusa-ku, Nagoya, 464-8601, Japan}



 \begin{abstract}
  Solar active region 12673 produced two successive X-class flares (X2.2 and X9.3) approximately 3 hours apart in September  2017. The X9.3 was the recorded largest solar flare in Solar Cycle 24. In this study we perform a data-constrained magnetohydrodynamic simulation taking into account the observed photospheric magnetic field to reveal the initiation and dynamics of the X2.2 and X9.3 flares. According to our simulation, the X2.2 flare is first triggered by magnetic reconnection at a local site where at the photosphere the negative polarity intrudes into the opposite-polarity region.  This magnetic reconnection expels the innermost field lines upward beneath which the magnetic flux rope is formed through continuous reconnection with external twisted field lines. Continuous magnetic reconnection after the X2.2 flare enhances the magnetic flux rope, which is lifted up and eventually erupts via the torus instability. This gives rise to the X9.3 flare. 
 \end{abstract}



% ==================================================================================================================================
    \section{Introduction} \label{sec:intro}
% ==================================================================================================================================
     Solar active region (AR) 12673 was observed in September 2017. The AR first appeared at the end of August and started growing on September 3, and went on to rapidly 
     strengthen its magnetic energy and helicity(\citealt{2019ApJ...872..182V}). During its evolution, AR12673 produced X-class flares several times in addition to C- and M-class 
     flares(\citealt{2021ApJ...908..132Y}). One of its solar flares, an X9.3 flare is the largest recorded solar flare in Solar Cycle 24, which generated  not only strong white-light 
     emission but also Gamma-Ray emission(\citealt{2019ApJ...877..145L}). It eventually produced a geo-effective coronal mass ejection(\citealt{2019SoPh..294..110W}, 
     \citealt{2020ApJS..247...21S}). In addition, an X2.2 flare was observed approximately 3 hours before the X9.3 flare(\citealt{2018ApJ...869...69M}, 
     \citealt{2018ApJ...856...79Y})(see also GOES plot shown in Fig. \ref{f1}(a)), {\it i.e.}, this AR produced successive X-class flares. AR12673 exhibited strong shearing and 
     converging flows(\citealt{2018ApJ...869...90W}) and also rotational motion(\citealt{2018ApJ...856...79Y}) at the flaring site, which are typical features of flare-productive 
     ARs(\citealt{2019LRSP...16....3T}). Furthermore, throughout the period of occurrence of the X2.2 and X9.3 flares, the negative polarity intruded, and eventually penetrated  
     into the opposites polarity region(\citealt{2017ApJ...849L..21Y}, \citealt{2019SoPh..294....4R}, \citealt{2020ApJ...894...29B}). From the multi-wavelength and the 
     spectroscopic data analysis, the intrusion has been considered to enhance the reconnection responsible for producing the successive X-class 
     flares(\citealt{2020ApJ...894...29B}).
    
    A nonlinear force-free field (NLFFF) extrapolation(\citealt{2012LRSP....9....5W}, \citealt{2016PEPS....3...19I}, \citealt{2017ScChD..60.1408G}) is a useful tool for inferring
    the structure of  three-dimensional (3D) magnetic field. The magnetic field {extrapolations} using the photospheric magnetic field as boundary condition have shown the 
    existence of the highly  twisted field lines before the X2.2 and X9.3 flares({\it e.g.,} \citealt{2018ApJ...856...79Y}, \citealt{2018ApJ...867...83I}, \citealt{2018ApJ...869...13J}, 
    \citealt{2018ApJ...867L...5L},  \citealt{2018ApJ...869...69M}, \citealt{2020ApJ...890...10Z}). The magnetic configuration has been found to be very complicated, for instance, 
    magnetic flux ropes (MFRs) with different sizes and shapes clustered on the main polarity inversion line (PIL) that produced the successive X-class flares and together 
    partly construct a double-decker MFR(\citealt{2018A&A...619A.100H}). The temporal variation in the coronal magnetic structure, particularly  the variation of the MFR, 
    has been discussed in relation to through the X2.2 and X9.3 flares(\citealt{2018ApJ...867L...5L}, \citealt{2019ApJ...870...97Z}). Because the X2.2 flare was confined, the 
    extrapolations showed highly twisted field lines even after the X2.2 flare while only the less twisted lines remained after the X9.3 flare. Furthermore a null point, which has been 
    suggested to play an important role in solar flares, was also found near the flaring site in the NLFFF extrapolations(\citealt{2018ApJ...869...69M}, \citealt{2019ApJ...870...97Z}), 
    while the intrusion was noticeable in the observation.
     
     Since the NLFFFs give information only on the static state of the magnetic field before and after the flare, we need to interpolate the flare dynamics by using them. In order to 
    reveal the flaring process without depending on the interpolation between the extrapolations, data-constrained/driven magnetohydrodynamic (MHD) simulations have been 
    performed(\citealt{2018ApJ...867...83I}, \citealt{2018ApJ...869...13J}, \citealt{2019A&A...628A.114P}). The data-constrained MHD simulation(\citealt{2018ApJ...867...83I}) used a 
    snapshot of the photospheric magnetic field as the boundary condition. In this case, they used the NLFFF reconstructed just before the X2.2 flare as the initial condition to a 
    time-dependent simulation. Because the NLFFF already contains free magnetic energy, we can expect it has the potential to produce a solar flare. The simulations 
    reproduced some observational aspects well(\citealt{2018ApJ...867...83I}, \citealt{2018ApJ...869...13J}) and have recently been used to investigate the particle motion 
    during the flare by using the MHD fields as the background field(\citealt{2020ApJ...902..147G}). On the other hand, the data-driven magnetofrictional 
    simulation(\citealt{2019A&A...628A.114P}) is driven by the time-dependent photospheric magnetic field, using the electric field derived from the time-series photospheric 
    magnetic field data(\citealt{2017SoPh..292..191L}). Therefore, the simulation starts much earlier than the flare occurrence time compared to the data-constrained MHD 
    simulation, thus it allows to reveal the energy accumulating process in the AR too. Note that  the results just imitates the observations because most of these simulations assume 
    simple solar atmosphere (for instance no chromosphere) and stay in the MHD framework without radiative transfer process. 
  
     In our previous paper, \cite{2018ApJ...867...83I} reported the simulation results for the eruption in which the multiple twisted lines are merged through the reconnection resulting 
    in the large erupting MFR and discussed the dynamics of the erupting MFR. On the other hand, the detailed initiation mechanism, the reason, and  mechanism of the successive 
    huge X-class flares are not fully understood.  In this study, we analyzed the simulation data together with observational data in detail to reveal the onset mechanism to produce 
    both the X2.2 and X9.3 flares. The rest of this paper is constructed as follows: the observations and numerical method are described in section 2; results and discussion are 
    presented in sections 3 and 4; and our conclusion is summarized in Section 5.     
    
% =====================================================================================================================================
     \section{Numerical Method $\&$ Observations} 
% =====================================================================================================================================
% -----------------------------------------------------------------------------------------------------------------------------------------------------------------------------------------------------------------------------------------        
     \subsection{Numerical Method}
% -----------------------------------------------------------------------------------------------------------------------------------------------------------------------------------------------------------------------------------------
    The numerical method is largely the same as the one described  in our previous study(\citealt{2018ApJ...867...83I}). In the following we shortly summarize its main 
    aspects. Both the NLFFF extrapolation method and the MHD simulation are based on the zero-beta MHD approximation given by the following equations:
% ------------------------------------------------------------------------
% Continuous Equation
% ------------------------------------------------------------------------
   \begin{equation}
   \rho(\vec{x}, t) =  |\vec{B}(\vec{x},t)|, 
   \label{con_hyp_eq}
   \end{equation}

% ------------------------------------------------------------------------
% Equation of Motion
% ------------------------------------------------------------------------
    \begin{equation}
    \frac{\partial \vec{v}}{\partial t} 
                         = - (\vec{v}\cdot \vec{\nabla})\vec{v}
                           + \frac{1}{\rho} \vec{J}\times\vec{B}
                           + \nu_i \vec{\nabla}^{2}\vec{v},
   \label{eq_of_mo}    
   \end{equation}

% ------------------------------------------------------------------------
% Induction equation
% ------------------------------------------------------------------------
  \begin{equation}
  \frac{\partial \vec{B}}{\partial t} 
                        =  \vec{\nabla}\times(\vec{v}\times\vec{B}
                        -  \eta_{\rm i}\vec{J})
                        -  \vec{\nabla}\phi, 
  \label{in_eq}
  \end{equation}

% ------------------------------------------------------------------------
% Ampere's low
% ------------------------------------------------------------------------
    \begin{equation}
    \vec{J} = \vec{\nabla}\times\vec{B},
    \label{Am_low}
    \end{equation}
  
% ------------------------------------------------------------------------
% Dedner 
% ------------------------------------------------------------------------
    \begin{equation}
    \frac{\partial \phi}{\partial t} + c^2_{\rm h}\vec{\nabla}\cdot\vec{B} 
      = -\frac{c^2_{\rm h}}{c^2_{\rm p}}\phi,
    \label{div_eq}
    \end{equation}
   where the formulation of viscosity and resistivity are different in the NLFFF and MHD computations, respectively. Therefore, the subscript ${\rm i}$ of 
   $\eta$ and $\nu$ corresponds to 'NLFFF' or 'MHD'. The length, magnetic field, density, velocity, time, and electric current density are normalized 
   by   
   $L^{*}$ = 244.8 (Mm),  
   $B^{*}$ = 0.25 (T), 
   $\rho^{*}$ = $|B^{*}|$ (kg/m$^{-3})$ ,
   $V_{\rm A}^{*}\equiv B^{*}/(\mu_{0}\rho^{*})^{1/2}$ (m/s),    
   where $\mu_0$ is the magnetic permeability,
   $\tau_{\rm A}^{*}\equiv L^{*}/V_{\rm A}^{*}$(s), and     
   $J^{*}=B^{*}/\mu_{0} L^{*}$(A),  
   respectively. $\phi$ is an artificial scalar potential used for controlling the errors  from $\vec{\nabla}\cdot \vec{B}$(\citealt{2002JCoPh.175..645D}), with the coefficients 
   $c_{\rm h}^2$, $c_{\rm p}^2$ in Equation (\ref{div_eq}) fixed to the constant values 0.04 and 0.1, respectively. The initial condition of density is given by 
   $\rho(t=0,\vec{x}) = |\vec{B}(t=0, \vec{x})|$ and the velocity is set at zero in all space in the NLFFF calculation and also MHD simulation.
  
  A numerical box of 244.8 $\times$ 158.39 $\times$ 195.84(M${\rm m}^3$) is , which is given as 1$\times$ 0.674 $\times$ 0.8 in its non-dimensional value, divided by 340 
  $\times$ 220 $\times$ 272 grid points. The data used in this study is yielded through a 2 $\times$ 2 $\times$ 2 binning process of the original SHARP data. The photospheric 
  magnetic field is preprocessed according to \cite{2006SoPh..233..215W}.  This process attempts to minimize L  which is sum of the total force $L_1$, the total torque 
  $L_2$,  the difference between the updated magnetic field and observed one $L_3$, and the smoothing $L_4$, as follows,
  \[
   L= \mu_{1}L_1 + \mu_{2}L_{2} + \mu_{3}L_{3}+\mu_{4}L_{4},
  \]
           
  \[
   L_1 = (\sum B_xB_z)^2 + (\sum B_yB_z)^2+(\sum B_z^2-B_x^2-B_y^2)^2,
  \]
  \[
   L_2 = (\sum x(B_z^2-B_x^2-B_y^2))^2 + (\sum y(B_z^2-B_x^2-B_y^2))^2+(\sum yB_xB_z-xB_yB_z)^2,
  \]
  \[
   L_3 = \sum (B_x-B_{x,obs})^2 + \sum (B_y-B_{y,obs})^2+\sum (B_z-B_{z,obs})^2,
  \]  
   \[
   L_4 = \sum (\Delta B_x^2 + \Delta B_y^2 + \Delta B_z^2 ).
  \]  
  
  In this study, $\mu_1$=$\mu_2$=1.0, $\mu_3$=$\mu_4$=$1.0\times 10^{-2}$ are used.
% =====================================================================================================================================
% NLFFF Extrapolation
% =====================================================================================================================================
    \subsubsection{Nonlinear Force-Free Field Extrapolation}  
    Details of the NLFFF extrapolation method are described in our previous studies(\citealt{2014ApJ...780..101I}, \citealt{2018ApJ...867...83I}, 
    \citealt{2018NatCo...9..174I}). Especially, the parameters used in this study are the same as those in our previous study(\citealt{2018ApJ...867...83I}). The viscosity and resistivity 
    are given as $\nu_{nlfff}=1.0 \times 10^{-3}$ and $\eta_{\rm nlfff} = \eta_0 + \eta_1 |\vec{J}\times\vec{B}||\vec{v}|^2/|\vec{B}|^2$, respectively, where $\eta_0 = 5.0\times 10^{-5}$ 
    and $\eta_1=1.0\times 10^{-3}$ in the non-dimensional unit. The second term is introduced to enhance reconnection which helps to make highly twisted lines observed in a central
    area of active regions. Furthermore, since the resistivity is proportional to the Lorentz force, it accelerates the relaxation of the non-equilibrium magnetic field 
    towards the force-free state.   
% ------------------------------------------------------------------------------------------------------------------------------------------------------------------------------------------------------------------------------------------
% Initial & Boundary Conditions
% ------------------------------------------------------------------------------------------------------------------------------------------------------------------------------------------------------------------------------------------
    The potential field is given as the initial condition, which is extrapolated using the Green function method(\citealt{1982SoPh...76..301S}). During the iteration, three components 
    of the magnetic field are fixed at each boundary, while the velocity is fixed to zero and the von Neumann condition $\partial /\partial n$=0 is imposed on $\phi$. Note that the bottom 
    boundary is fixed according to 
   \[
    \vec{B}_{\rm bc} = \zeta \vec{B}_{\rm obs} + (1-\zeta) \vec{B}_{\rm pot},
   \]
   where $\vec{B}_{\rm bc}$  is the transverse component determined by a linear combination of the observed magnetic field ($\vec{B}_{\rm obs}$), and the potential magnetic 
   field($\vec{B}_{\rm pot}$). $\zeta$ is a coefficient in the range of 0 to 1. When $R=\int |\vec{J}\times\vec{B}|^2$dV, which is calculated over the interior of the computational domain, 
   falls below a critical value denoted by $R_{\rm min}$, during the iteration, the value of the parameter $\zeta$ is increased to $\zeta = \zeta + {\rm d}\zeta$. In this paper, 
   $R_{\rm min}$ and d$\zeta$ have the values $1.0 \times 10^{-2}$ and 0.02, respectively. If $\zeta$ is equal to 1, $\vec{B}_{\rm bc}$ is completely consistent with the observed data.  
   Furthermore, the velocity is controlled as follows. If the value of $v^{*}(=|\vec{v}|/|\vec{v_{A}}|)$ is larger than $v_{\rm max}$(here set to 0.02), then the velocity is modified as 
   follows: $\vec{v} \Rightarrow (v_{\rm max}/ v^{*}) \vec{v}$. These processes would help avoid a sudden jump from the boundary into the domain during the iterations.
   
   The NLFFFs are reconstructed from the photospheric magnetic field obtained at 02:36 UT and 08:36 UT on September 6. Figure \ref{f1}(b) shows the 3D magnetic field of the 
   NLFFF at 08:36 UT. The strongly twisted lines have strong current density while the overlying field lines take a state close to the potential field. The comparison with extreme 
   ultraviolet image and how much the magnetic field satisfies the force-free state are described in Appendix in \cite{2020ApJ...894...29B}. 
      
% ============================================================================================================
% MHD Simulation
% ============================================================================================================
    \subsubsection{Magnetohydrodynamic Simulation}
    The zero-beta MHD simulations are  carried out using the NLFFFs as the initial conditions which are reconstructed at 08:36UT and 02:36UT 
    on September 6, respectively, hereafter denoted Run A and B.  The resistivity and viscosity are set to a uniform $\eta_{\rm mhd}=1.0\times 10^{-5}$ 
    and $\nu_{\rm mhd}=1.0 \times 10^{-4}$. The density is initially given as $\rho(\vec{x}, t=0)=|\vec{B}(\vec{x}, t=0)|$, even during the MHD simulation, 
    we imposed $\rho(\vec{x}, t)=|\vec{B}(\vec{x}, t)|$; however, the dynamics in the lower corona as focused in this study are not so different from those 
    obtained using a mass conservation law equation(\citealt{1999ApJ...518L..57A}, \citealt{2014ApJ...788..182I}). Note that there is no physical 
    significance of the density obtained in this simulation. The density  is fixed at boundaries during the calculation. The normal component $B_n$ 
    is fixed at all boundaries but the transverse component, $\vec{B}_{\rm t}$, is derived from an induction equation under the velocity equals to zero and $\eta$ 
    is also fixed to zero at each boundary. For instance, at the bottom surface, the time variations of $B_x$ and $B_y$ are derived from 
    \begin{equation}
    \frac{\partial B_x}{\partial t}\mid_{z=0} =  \frac{\partial E_y}{\partial z} - \frac {\partial \phi}{\partial x}, 
    \label{mag_bc_x}
    \end{equation}
    \begin{equation}
    \frac{\partial B_y}{\partial t}\mid_{z=0} =  -\frac{\partial E_x}{\partial z} - \frac{\partial \phi}{\partial y}, 
    \label{mag_bc_y}
    \end{equation}
    where $E_x = (\eta_{\rm mhd} \vec{J} - \vec{v} \times \vec{B})_{x}$ and  $E_y = (\eta_{\rm mhd} \vec{J} - \vec{v} \times \vec{B})_{y}$ are 
    x and y components of the electric field.  Boundary conditions of the other variables are identical to those in the NLFFF calculation.   
 
 % ===================================================================================================================================
 % Observations
 % ===================================================================================================================================
     \subsection{Observations}
     To reconstruct the NLFFFs,  we use the photospheric magnetic field obtained at 02:36UT and 08:36 UT which are 6hours 20 minutes and 20 minutes before the X2.2 flare, 
     respectively,  on September 6 taken by the Helioseismic and Magnetic Imager (HM: \citealt{2012SoPh..275..207S}) onboard the Solar Dynamics 
     Observatory(SDO :\citealt{2012SoPh..275....3P}). This observation is shown in the space-weather HMI active region patch format(\citealt{2014SoPh..289.3549B}). 
     Far-ultraviolet(FUV) 1600 \AA \ image taken by  the  Atmospheric Imaging Assembly(AIA: \citealt{2012SoPh..275...17L}) onboard the SDO is also used to compare with the 
     NLFFF and dynamics obtained from the simulation. 
     
\begin{figure}
  \epsscale{1.}
  \plotone{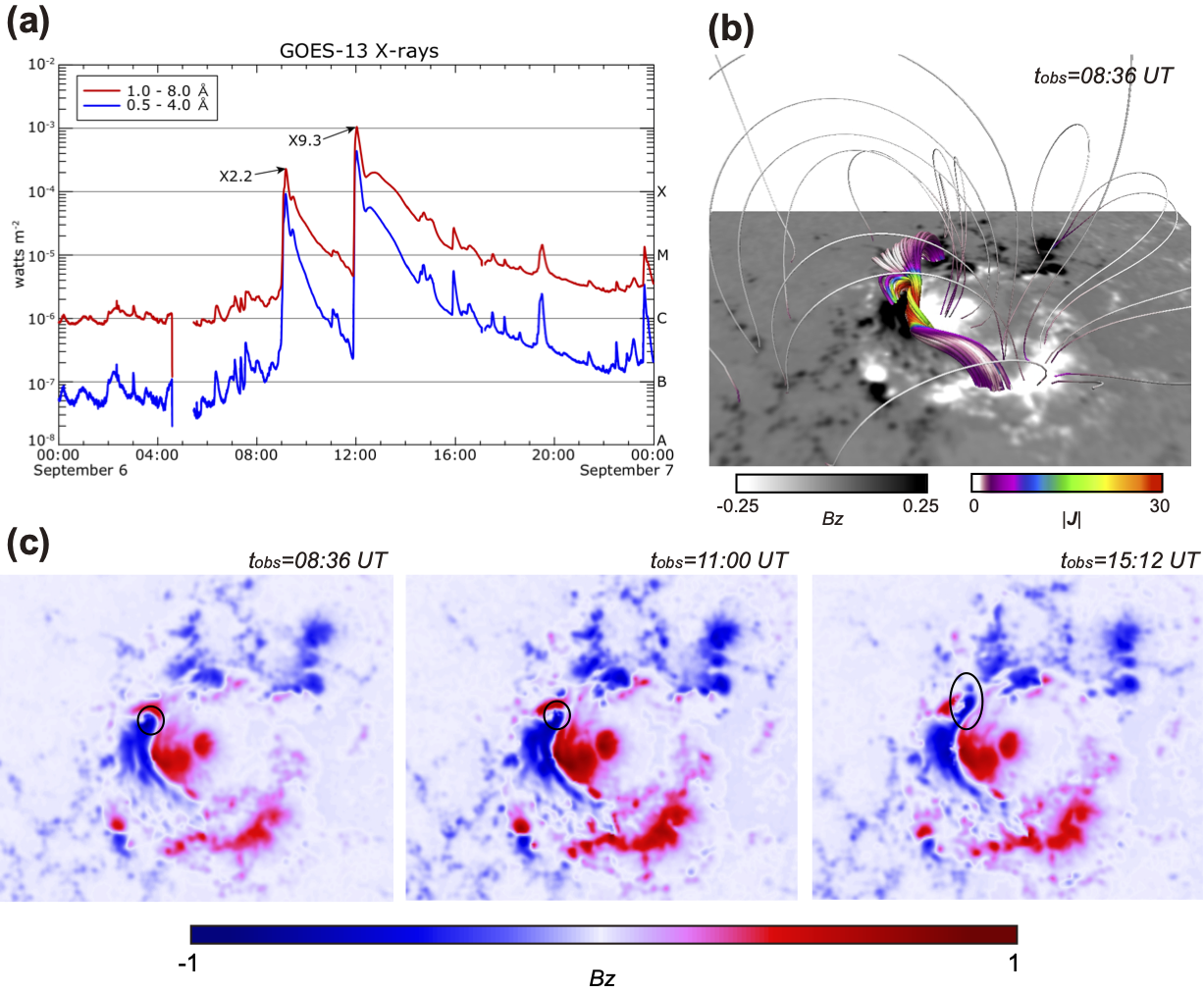}
  \caption{
               (a) Time profile of the X-ray flux measured by the GOES-13 satellite on September 6. The X2.2 and X9.3 flares  indicated by the arrows started at 08:57 UT and 11:53 UT, 
                     respectively. $0.5-4.0$ \AA \ passband are plotted, respectively, and  the solar X-ray outputs in the 1$-$8 \AA \. 
               (b) Three-dimensional magnetic field reconstructed under an NLFFF approximation based on the photospheric magnetic field. The photospheric magnetic field is obtained 
                     at 08:36 UT on September 6 which is 20 minutes before the X2.2 flare. The color of field lines indicates the strength of the current density $|\vec{J}|$.
               (c) Photospheric magnetic field (normal component) is shown at $t_{obs}=$08:36 UT, $t_{obs}=$11:00 UT and $t_{obs}=$15:12 UT, respectively. The panels cover
                    the time from before the X2.2 flare observed at 08:57 UT to after the X9.3 flare observed at 11:53 UT. Throughout the X2.2 and X9.3 flares, the negative magnetic 
                    field indicated by the black circles intrudes into the positive polarity region and eventually penetrates.
               }
  \label{f1}
  \end{figure}

\begin{figure}
    \epsscale{.9}
    \plotone{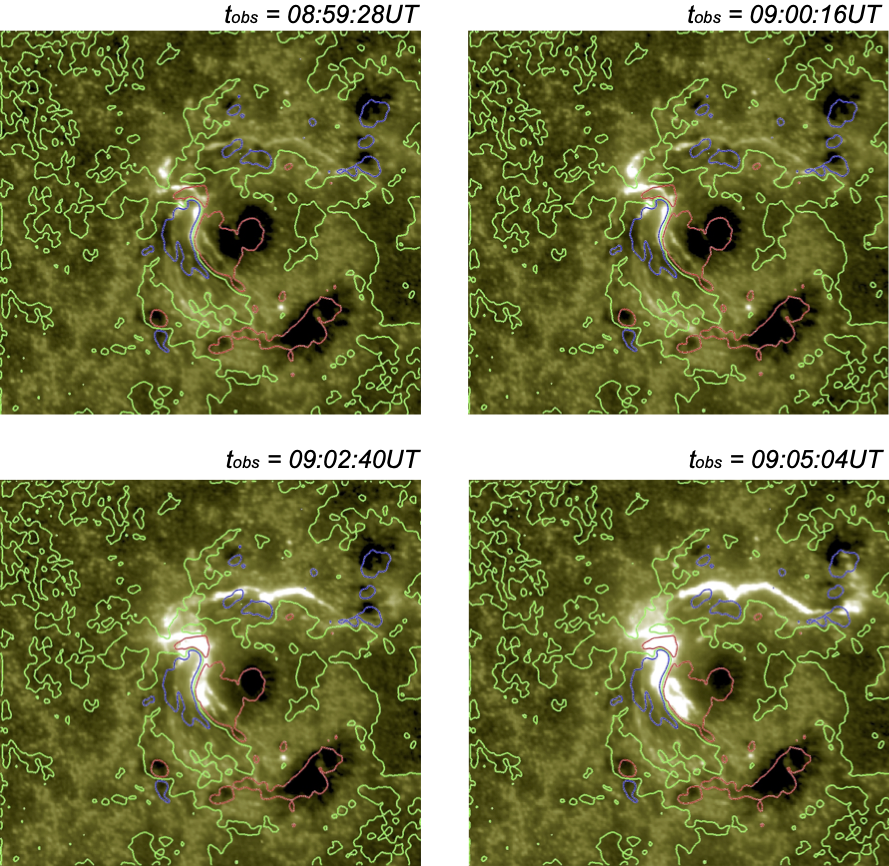}
      \caption{
                   The images are obtained from AIA 1600 \AA \ during the X2.2 flare. The red and blue lines correspond to the contours of $B_z=$0.25 and 
                   $B_z=$-0.25, respectively. The green lines represent the polarity inversion lines.} 
   \label{f2}
   \end{figure}

\begin{figure}
    \epsscale{.9}
    \plotone{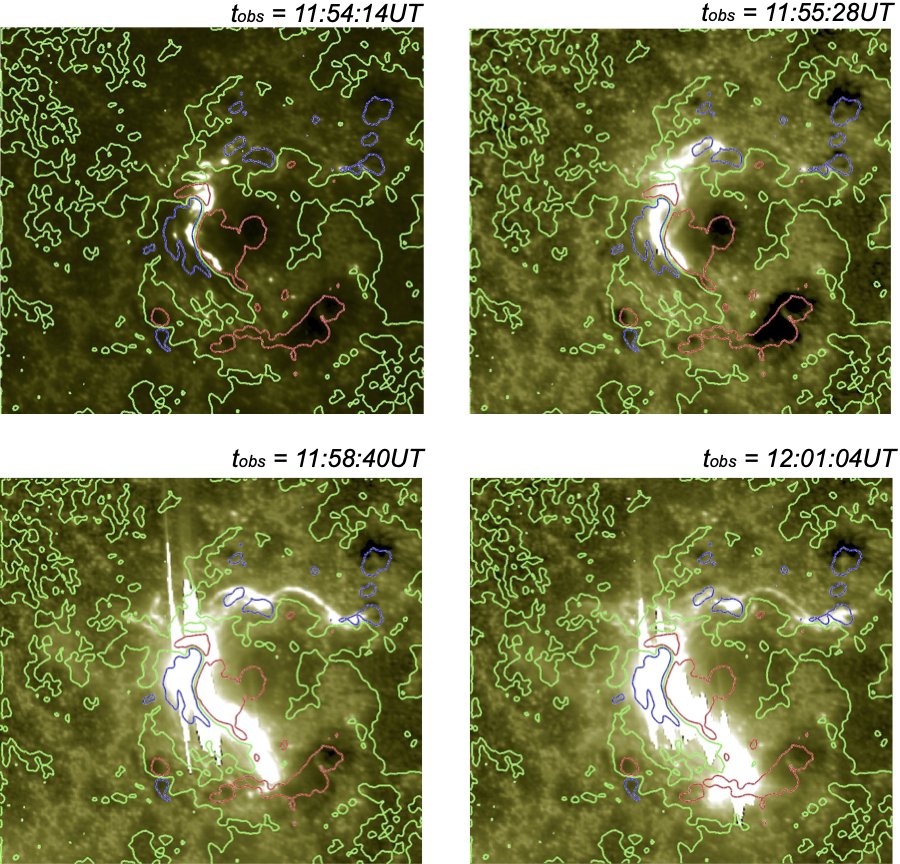}
    \caption{The images are taken from AIA 1600 \AA \ during the X9.3 flare. The format is the same as Figure \ref{f2}.} 
   \label{f3}
   \end{figure}

     As outstanding features, the negative polarity intrudes and eventually penetrates into the positive polarity during the successive flares as shown in Fig. \ref{f1}(c). 
     For instance, \cite{2020ApJ...894...29B} suggests that the reconnection is enhanced there, which would be important to cause the successive flares. Figures.\ref{f2} and \ref{f3} 
     shows the AIA 1600 images obtained during the X2.2 and X9.3 flares. In both cases, strong brightening is observed at the PIL where the strong twisted lines exist as see in 
     Fig.\ref{f1}(b). Although both profiles are similar, the X9.3 flare shows stronger enhancement and the brightening further extends to the southward along the PIL. 
     
 % ----------------------------------------------------------------------
% 3D Field Lines and those dynamics
% ----------------------------------------------------------------------

% ---------------------------------------------------------------------
% AIA Image during the X2.2 flare
% ---------------------------------------------------------------------
   \clearpage
   
% ---------------------------------------------------------------------
% AIA Image during the X9.3 Flare 
% ---------------------------------------------------------------------
   \clearpage

% ===========================================================================================================================
    \section{Results} \label{sec:floats}
% ===========================================================================================================================
    \subsection{Initiation of the X2.2 Flare}
    Figure \ref{f4}a shows an FUV 1600 \AA \ image taken by  AIA at the time that corresponds to the onset of the X2.2 flare.  In Fig.\ref{f4}(b), a selection of field lines traced 
    from the MHD simulation at $t=$0.28 when the reconnection starts is superimposed on this image, from which their footpoints are anchored well to the strong brightening 
    sites(see also the inset in Fig.\ref{f4}(b)). Therefore those field lines induce the X2.2 flare. Figure \ref{f4}(c) shows the side view of the field lines where the color of the line 
    corresponds to the value of  $|\vec{J}|/|\vec{B}|$. We found the strong current density indicated by the circle to be located directly above the region where the negative 
    polarity intrudes into the opposite polarity region. In the very early stage of the simulation (Figs.\ref{f4}(c)-(d)), magnetic reconnection occurs, resulting in new twisted 
    lines. Note that since the horizontal magnetic field at the photosphere changes according to equations (\ref{mag_bc_x}) and (\ref{mag_bc_y}) in the MHD simulation, the 
    magnetic field gradually deviates from the NLFFF at $t$=0 following which the magnetic field above changes because the NLFFF cannot take a perfect force-free state 
    which causes the residual force to act on the magnetic field. In addition the velocity limit imposed in the NLFFF calculation is deactivated, and consequently, 
    reconnection is excited in the MHD simulation while the NLFFF keeps the configuration as shown in Fig.\ref{f1}(b).
        
    The vertical cross-section shown in panel (c) and (d) in Fig.\ref{f4} draws the decay index $n$. The decay index(\citealt{1978mit..book.....B}, \citealt{2006PhRvL..96y5002K}) 
    is defined as
    \begin{equation}
     n= -\frac{z}{B_{\rm p_t}}\frac{\partial B_{\rm p_t}}{\partial z},
    \end{equation}     
    where $B_{p_t}$ denotes the horizontal component of the external field surrounding the MFR. Here, we assume a potential field as the external field. This value is a proxy 
    of the torus instability(TI:\citealt{2006PhRvL..96y5002K}). When the axis of the MFR reaches the region in red where the condition $n \ge 1.5$ is satisfied, TI works on the 
    MFR {\it i.e.}, the upward hoop force due to the current flowing inside the MFR is dominant over the strapping force from the external field. Note that the threshold strongly 
    depends on the shape or boundary condition of the MFR(\citealt{2010ApJ...718..433O}, \citealt{2015ApJ...814..126Z}) and mass sustained by the 
    MFR(\citealt{2019ApJ...873...49J}).  We found that the twisted lines formed through the reconnection cannot reach the region where those become unstable to the TI. 
    Therefore, we suggest that the MFR at this time is stable to the TI.
    
    We already discussed the possibility of the onset driven by kink instability(KI) in \cite{2018ApJ...867...83I}.  According to their results, the highly twisted lines which 
    satisfy the KI(\citealt{2004A&A...413L..27T}) are not found in the NLFFF (See also Figure \ref{f9} in this paper), and the possibility of KI is low even in highly twisted 
    erupting MFR created through the reconnection.  Therefore, we rule out KI as an initiation mechanism for this event.
    
    Figure \ref{f4}(e) shows the AIA 94 \AA \ image when the X2.2 flare just starts. The strong brightening area is observed along with the PIL marked by yellow dashed 
    circle. The field lines are plotted over the AIA 94 \AA \ image in Figure \ref{f4}(f) where the color is in the same format as the one in Figs. \ref{f4}(c) and (d)  
    and the reconnection starts at t=0.28 in the simulation at same area. Therefore, the observation supports that the initiation is triggered by the reconnection 
    at the strong current region formed by the intruding motion of the sunspot.
  
 % -------------------------------------------------------------------------------------------------------------------------------------------------------------------------------------------------------------------------  
      \subsection{Three-dimensional Dynamics of the Magnetic Field}    
 % --------------------------------------------------------------------------------------------------------------------------------------------------------------------------------------------------------------------------
    Figure{\ref{f5}}(a) plots the field lines causing the X2.2 flare again, as well as another set of twisted lines that do not participate in the onset of the X2.2 flare. The X2.2 
    flare is triggered by reconnection in the local site indicated by the square, and generates long twisted lines as shown in Fig.\ref{f5}(b). This reconnection plays a role 
    in expelling a part of the twisted lines which exists along the PIL. As a result, another set of twisted lines, which exists at the outside, starts a reconnecting under the long 
    twisted lines which are formed in an early phase of the X2.2 flare, as shown in Fig. \ref{f5}(b) . The inset of the Fig. \ref{f5}(b) shows an enlarged view of the newly created 
    MFR indicated  by the arrow. Continuous reconnection takes place below the MFR, which supplies twist and makes the MFR bigger(later discussed), eventually causing 
    the eruption that triggers the X9.3 flare(Figs.\ref{f5}(c)-(d)). Figure \ref{f2} shows that, after the initial brightening at $t_{obs}$=09:00UT, the brightening is further enhanced 
    around the PIL. Therefore the reconnection still occurs around the PIL. The simulation in this study proves that the reconnection is caused by the sheared field lines under 
    the  twisted lines created at the onset stage of the X2.2 flare. This reconnection generates a fundamental MFR as shown in the inset of Fig.\ref{f5}(b), 
    which evolves into the large MFR via the subsequent reconnection and leads to the X9.3 flare. 
      
    We compare the erupting MFR  and current density $|\vec{J}|$(or $|\vec{J}|/|\vec{B}|$) with the AIA 1600 \AA \ image obtained during the X9.3 flare. Figure \ref{f6}(a) shows 
    an erupting MFR  with the AIA 1600 \AA \ image projected to the simulation bottom which is taken at 12:01:04 UT. The observational time $t_{obs}$=12:01:04 UT of AIA 
    1600 \AA \ is just after the strongest brightening of the X9.3 flare which we interpret as the time at which the MFR goes up away from the active region. Therefore, we 
    select the last time of the simulation for the comparison. By comparing between the AIA images obtained during the X2.2 and the X9.3 flare shown in Figs. \ref{f2} and \ref{f3},  
    we find that the brightening related to the X9.3 flare extends further to the south along the PIL compared to the case for the X2.2 flare. In this simulation, one leg of the 
    erupting MFR is anchored to the extended brightening region. Figure \ref{f6}(b) focuses on the lower area beneath the erupting MFR. The current sheet is formed above 
    post-flare loops whose footpoints are anchored to the intense brightening region shown in the AIA image. Since a bifurcation of up- and down-flows are {present close to} 
    the current sheet, we suggest that the reconnection takes place there.  Figure \ref{f6}(c) plots field lines which are traced in the y-z plane to facilitate the description of the 
    dynamics by reducing into the two-dimensional space. This figure clearly shows that the current sheet is surrounded by antiparallel field lines which is consistent with the 
    classical standard flare model({\it e.g.} see \citealt{2002A&ARv..10..313P} and it's 3D generalization see, {\it e.g.}, 
    \citealt{2012A&A...543A.110A}, \citealt{2013A&A...555A..77J}). Figure \ref{f6}(d) plots the volume rendering of current density $|\vec{J}|$ on the AIA 1600 \AA \ image. The 
    distribution of the $\vec{|J|}$ precisely captures the brightening area enhanced in the AIA 1600 \AA \ image. Therefore, our simulation would produce the MHD process of both 
    the X2.2 flare and X9.3 flares. On the other hand,  the brightening which appears on the northern side is not reproduced in our simulation. One of the reasons that 
    the initial NLFFF does not reproduce the magnetic field on the northern area {is because} the magnetic field is weaker than that of the central 
    area(\citealt{2020ApJ...894...20L}).
    
     From these results, we concluded that the X2.2 flare is driven by the reconnection which takes place at the location indicated by the dashed circle in Fig. \ref{f4}(c), 
     while the reconnection drives the X9.3 flare at the strong current region that is located above the PIL as shown in Fig.\ref{f6}(b). Since the size of the regions where the 
     energy is released through the magnetic reconnection is much different between the first and second flares, our simulation is therefore consistent with the fact that the 
     second flare was bigger than the first one.

% -------------------------------------------------------------------------------------------------------------------------------------------------------------------------------------------------------------------------------
     \subsection{Initiation of the X9.3 Flare}
% -------------------------------------------------------------------------------------------------------------------------------------------------------------------------------------------------------------------------------
     \subsubsection{Tempral evolution of Height and Velocity of the Erupting MFR}
    To find out the start of the X9.3 flare, the initiation of the eruption of the large MFR should be discussed due to the association of the flare with the eruption of the MFR. 
    We trace the temporal evolution of the field line in black shown in Fig.\ref{f7}(a), which is one of the components of the erupting MFR, to understand the transition from pre- 
    to post-eruption of the MFR. Figure \ref{f7}(b) shows the temporal evolution of the height of the erupting MFR in red, and the magnetic flux where the magnetic twist($T_w$) 
    of  the field lines satisfies $T_w \le -1.0$ in blue, respectively. The magnetic twist(\citealt{2006JPhA...39.8321B}) is defined as  
    \begin{equation}
    T_{\rm w}=\frac{1}{4\pi} \int \frac{\vec{\nabla}\times\vec{B}\cdot\vec{B}}{|\vec{B}|^2}dl,
    \end{equation}
    where $dl$ is a line element of a field line. Note that, in this study, the twist is calculated under the condition $|B_z|>2.5\times10^{-3}T$ on the surface. Since $T_w$ 
    measures the number of turns of two infinitesimally close field lines, the twist is strictly different from the winding number of the field line around the magnetic axis of the 
    MFR(\citealt{2016ApJ...818..148L}, \citealt{2018SoPh..293...98T}). In figures \ref{f7}(b) and (c) {$t \sim 0$ corresponds to the onset time of the X2.2 flare. The height profile 
    shows a slow-rising phase, $t \le 2 $, and a fast raising phase, $t \ge 2$, both of which are typical profiles of the erupting MFR. From Figures \ref{f7}(b) and (c) we can thus 
    deduce that the X9.3 flare starts around $t \sim 2$ when the MFR shifts to the fast-rising mode. During the slow-rising phase, the size of the MFR, which is occupied with 
    $T_w \le -1.0$, increases, {\it i.e.}, reconnection constantly takes place after the X2.2 flare. Therefore, a pre-eruption MFR is produced through the X2.2 flare and by the 
    continuous reconnection in the slow-rising phase, until the MFR is accelerated. 
     
     We also attempt to answer the fundamental question of what mechanism drives the rapid acceleration of the MFR. Figure \ref{f7}(c) plots the temporal evolution of 
    velocity of the erupting MFR in blue, in addition to the height profile in red, which is the same as the one shown in Fig. \ref{f7}(b). The velocity is plotted on a log scale, in 
    which it grows linearly during $t=1.5 \sim 2.5$. Since this period covers the bifurcation from the slow- to fast-rising phase, an instability drives the eruption. Note that after 
    $t \approx 4$, the velocity gradually decreases since the side boundary suppresses the dynamics of the MFR. Figures \ref{f8}(a)-(c) present the snapshots of the 3D field 
    lines from the end of the slow-rising phase to the start of the fast rising phase ({\it i.e.,} initiation of the MFR), respectively, with the decay index plotted on the vertical 
    cross-section. At $t=1.4$, a part of the flux rope enters the region where the instability would be triggered in idealized conditions($n \ge$1.5). At $t=$2.54, further twisted 
    lines, which are enhanced by the reconnection, enter the region of instability. Therefore, this result suggests that the X9.3 flare is driven by the torus instability, which is 
    consistent with results analyzed by the NLFFF extrapolations performed by \cite{2020ApJ...890...10Z}.

 % -----------------------------------------------------------------------------------------------------
    \subsubsection{Comparison of an Evolution of the MFR with AIA images}
 % -----------------------------------------------------------------------------------------------------
     We compare the evolution of the magnetic field at the time of the initiation of the X9.3 flare with SDO/AIA observations. The bottom in Fig. \ref{f8}(d) is set at the 
     AIA 1600\AA \ mage taken at the very early stage of the X9.3 flare. These results suggest that the brightening in AIA 1600\AA \ is enhanced at the time corresponding to 
     when the erupting MFR is formed through the reconnection and gradually arises because the field lines whose footpoints anchored to the brightening areas are changing 
     to the long field lines through the reconnection. Afterward, as seen in Fig.\ref{f3}, the FUV brightening is strongly enhanced and extended southward. In this simulation, 
     the erupting MFR, named "MFR1", in Fig.\ref{f8}(d), reconnects with the twisted field lines named "TFL" during the eruption. Consequently, the large MFR is formed as 
     shown in Fig.\ref{f6}(a), and one  footpoint of the newly created MFR is anchored to the brightening region extending southward, as seen in AIA 1600\AA. 
     Therefore, this simulation offers a sound explanation of the behavior of  the 1600\AA \ images observed in the X9.3 flare.

% ----------------------------------------------------------------------
% 3D Field Lines and those dynamics
% ----------------------------------------------------------------------
  \begin{figure}
  \epsscale{.8}
  \plotone{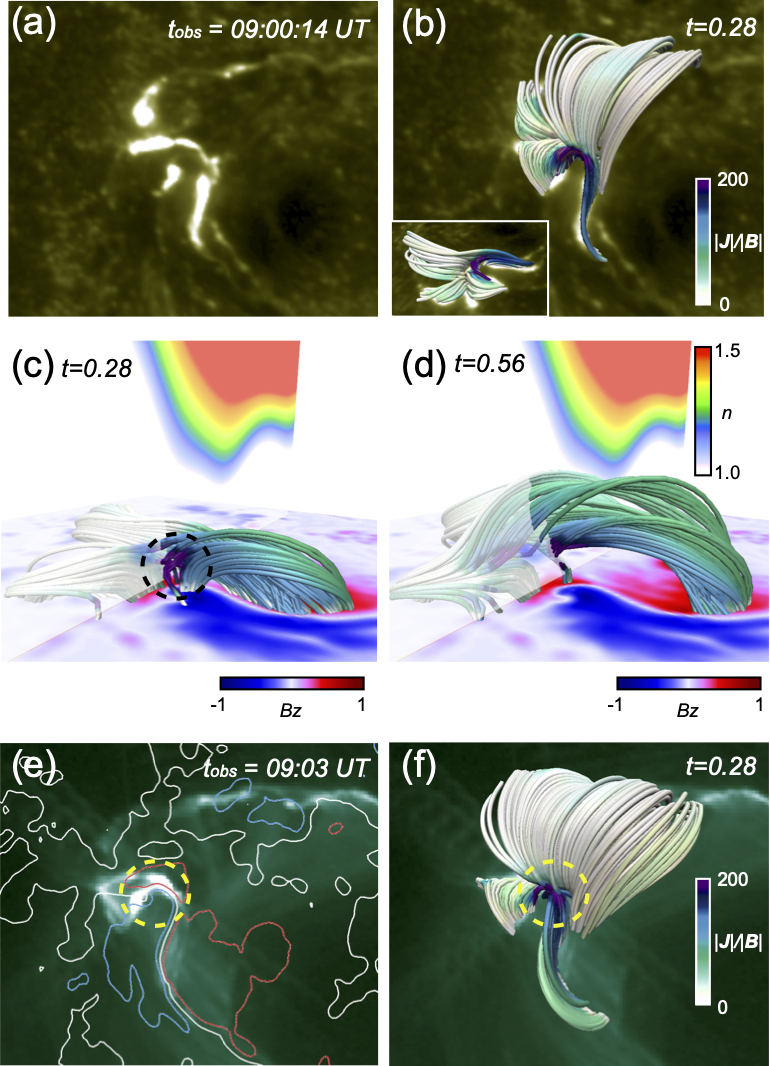}
  \caption{
                (a) AIA 1600 \AA \ image at 09:00 UT at the onset  phase of the X2.2 flare.
                (b) Field lines obtained from the MHD simulation at t=0.28, when the reconnection starts, are superimposed on the AIA image for the run where the initial 
                     condition given by the NLFFF is reconstructed at 08:36 UT. The color of the field lines correspond to the value of $|\vec{J}|/|\vec{B}|$. The small inset is 
                     the side view. 
           (c)-(d) The onset of the X2.2 flare is inferred from the simulation. The field lines are plotted obtained from the MHD simulation at $t=0.28$ and $t=0.56$, respectively. 
                      These field lines would induce the X2.2 flare. The bottom panel shows the $B_z$ distribution, and the decay index value is painted on the vertical cross section 
                      in a range from 1.0 to 1.5.
                (e) AIA 94 \AA \ image at 09:03 UT at the onset phase of the X2.2 flare with contours of $B_z=\pm{0.25}$ and PIL in red(+), blue(-) and white, respectively.
                     Strong brightening area is marked by the yellow dashed circle.
                (f) The field lines are plotted over the AIA 94 \AA \ image where the color corresponds to $|\vec{J}|/|\vec{B}|$. The yellow dashed circle is same to the one in (e).}
  \label{f4}
  \end{figure}
    
 % ------------------------------------------------------------------------------
% Temporal Evolution of Flux Tube Velocity and Decay Index
% ------------------------------------------------------------------------------
  \begin{figure}
  \epsscale{.8}
  \plotone{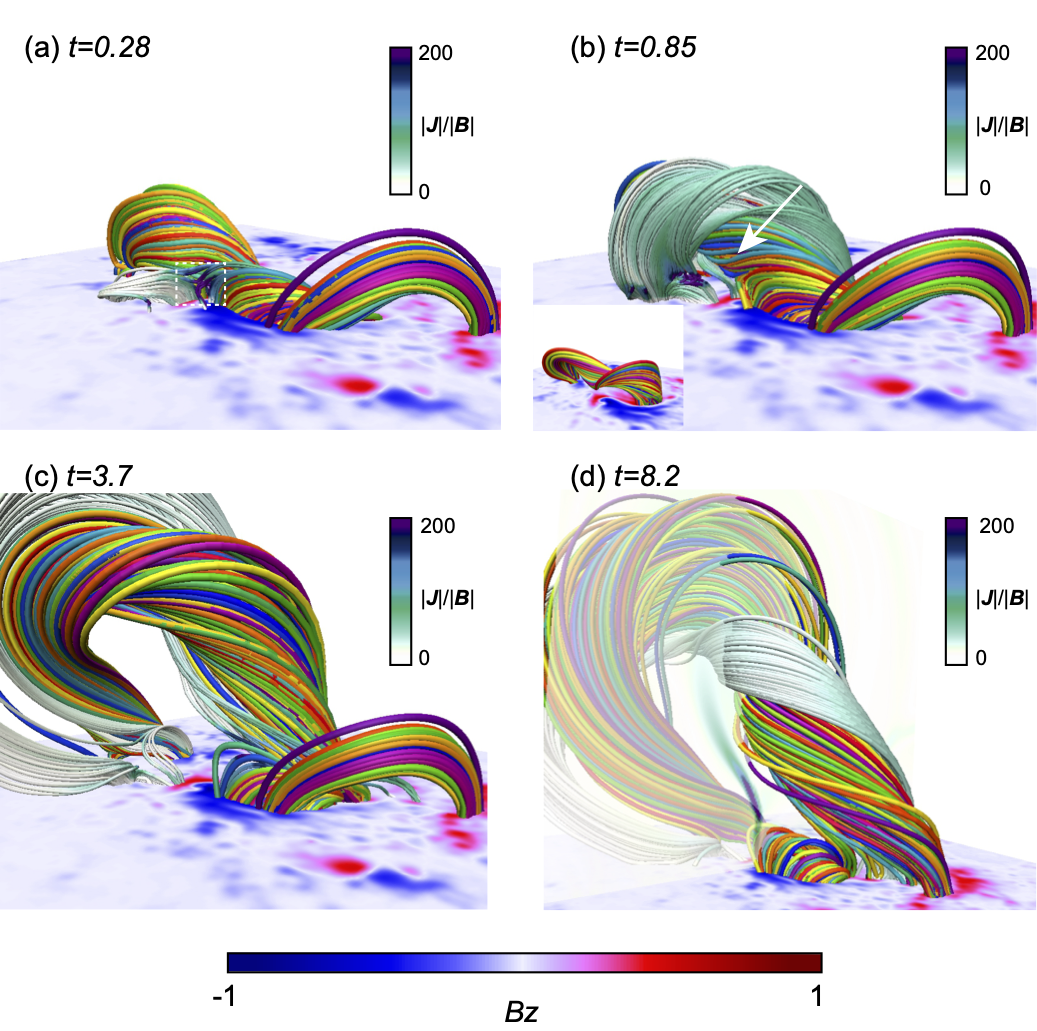}
  \caption{ 
               (a)-(d) The temporal evolution of magnetic field lines causing the X2.2 and X9.3 flares. The field lines highlighted by the value of $|\vec{J}|/|\vec{B}|$ correspond to 
                          those that are related to the onset of the X2.2 flare. The format is the same as Figure \ref{f1}. Another set of field lines are shown that belong to a large 
                          erupting MFR inducing  the X9.3 flare.  The vertical cross section in (d) is plotted by $|\vec{J}|/|\vec{B}|$  distribution. The arrow shown in (b) points toward 
                          the newly created MFR, whose enlarged view is in the inset.}
  \label{f5}
  \end{figure}

% -------------------------------------------------------------------------
% Top View of 3D Magnetic Structure
% -------------------------------------------------------------------------
\begin{figure}
  \epsscale{1.}
  \plotone{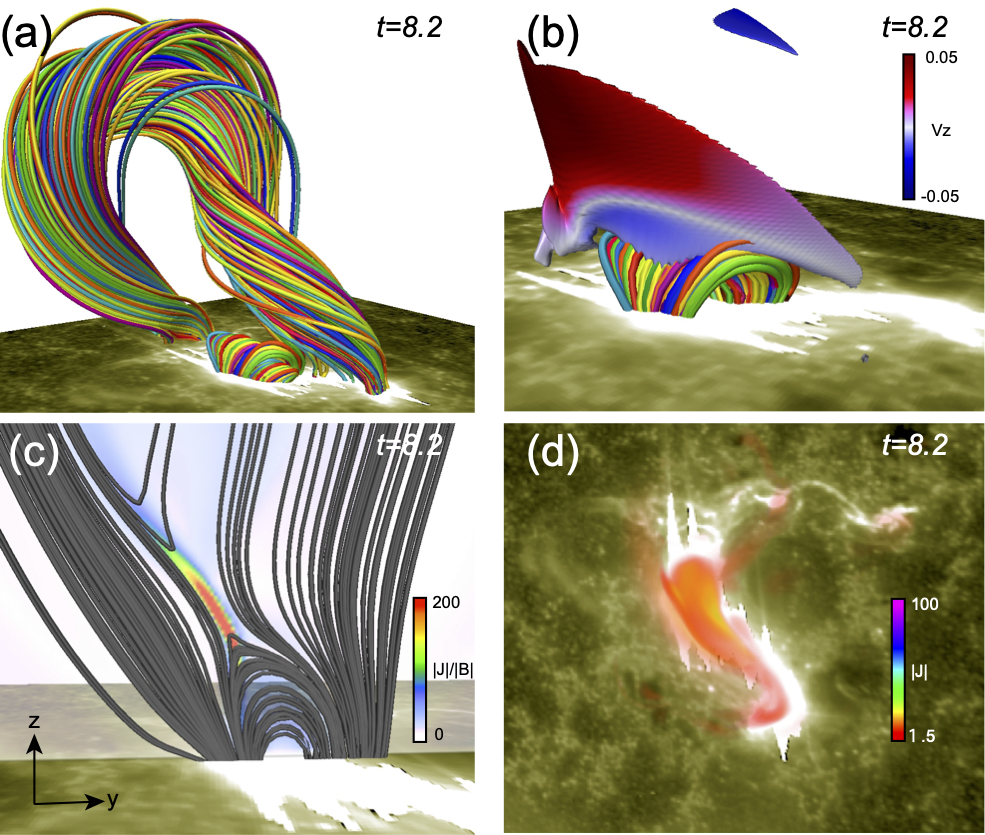}
  \caption{
                (a) The field lines constituting the erupting MFR are plotted where the bottom is set as the AIA 1600 \AA \ image taken at $t_{obs}=$12:01:04 UT on September 6 2017. 
                (b) The enlarged view close to the surface under the erupting MFR where the AIA 1600 \AA image set at the bottom is same as in (a). 
                     The surface corresponds to the iso-surface of $|\vec{J}|/|\vec{B}|$= 80 where the color represents $v_z$. 
                (c) The field lines in gray are traced in y-z plane where the $|\vec{J}|/|\vec{B}|$ is plotted on the vertical cross section.
                (d)  The volume rendering of $|\vec{J}|$ is superimposed on the AIA 1600 \AA \ image.
                }
  \label{f6}
  \end{figure}
  
 % -------------------------------------------------------------------------  
% Quantitative value of eruptive flux tube          
% -------------------------------------------------------------------------   
  \begin{figure}
  \epsscale{1.}
  \plotone{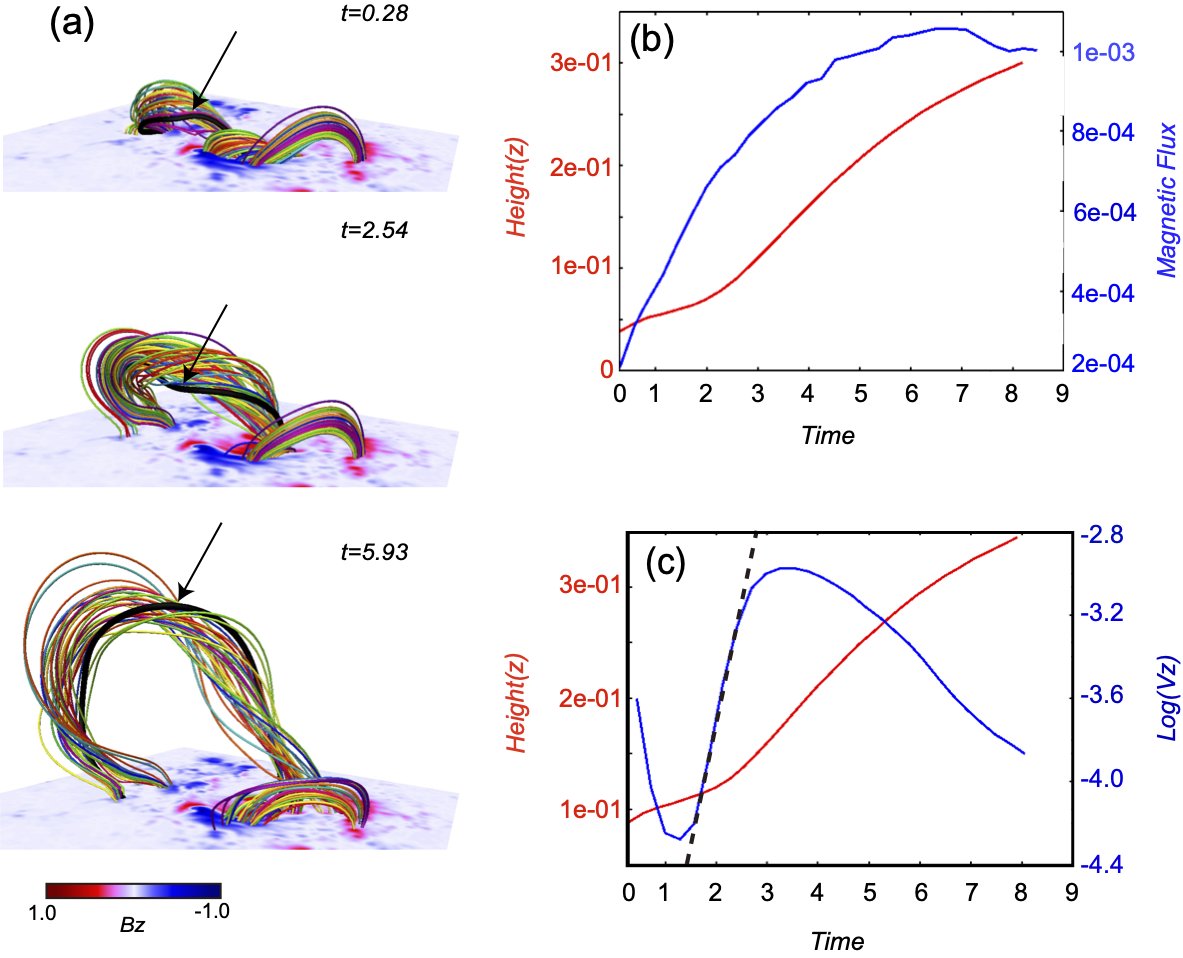}
  \caption{
               (a) Temporal evolution of the erupting MFR. We temporally trace the black line (indicated by black arrow), which is one of components of the erupting MFR.
                    We put a particle at the point $\vec{r}=(0.392,0.35,0.039)$ at t = 0.28 and trace the particle during the simulation.
               (b) Temporal evolution of the height of the erupting MFR denoted by red and the magnetic flux that satisfies  $T_w \le -1.0$ denoted by blue.
               (c) Temporal evolution of velocity of the MFR in blue added to the height profile in red. The dashed line represents the exponential growth phase of the 
                     velocity.
                }
  \label{f7}
  \end{figure}
     
% -------------------------------------------------------------------------  
% Variation of the magnetic fields through an anomalous resistivity
% -------------------------------------------------------------------------   
  \begin{figure}
  \epsscale{.8}
  \plotone{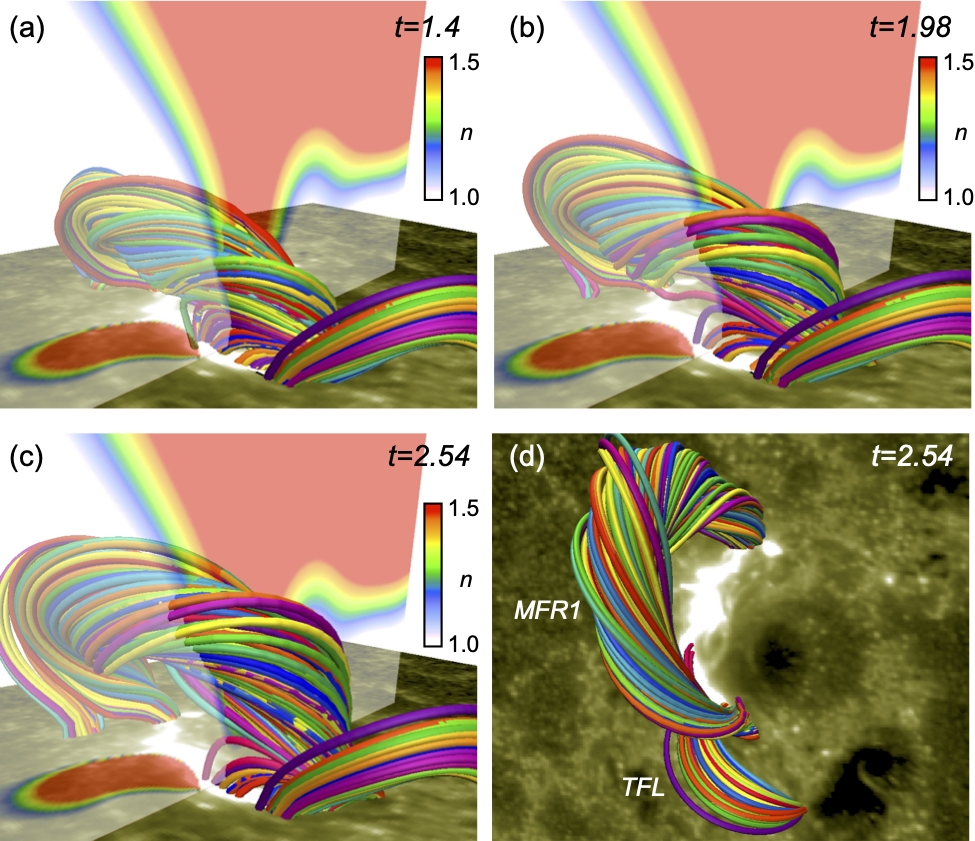}
  \caption{
               (a)-(c) Evolution of the erupting MFR from $t$=1.4-2.54 and the distribution of the decay index derived from the potential field. In this period, the speed 
                          of the rise of the MFR switches from slow to fast. The bottom corresponds to the AIA 1600 \AA \ image taken at $t_{obs}$=11:54:28 UT, which corresponds 
                          to the early stage of the X9.3 flare.                  
                    (d) The top view of (c).}
  \label{f8}
  \end{figure}
  \clearpage
  
% ===================================================================================================================
   \section{Discussion}  
% ====================================================================================================================
    \subsection{A Possibility of Another Scenario for the Solar Eruption}
    \subsubsection{Another Candidate Triggering Solar Eruption}
    In the previous sections, we have demonstrated that the reconnection occurring at a local area, where the negative polarity intrudes into the opposite one 
    (see Fig.\ref{f1}(c)), is important to produce the X2.2 flare. Further continuous reconnection drives the evolution of the MFR and eventually causes  the 
    X9.3 flare. We now discuss another scenario to produce the successive flares. Since AR12673 shows a very complex magnetic field distribution, 
    magnetic null points are likely to exist at several points. For example, Fig. \ref{f9}(a) and (b) (top views in Fig.\ref{f9}(c) and (d)) show the location of the 
    null points in purple field lines and yellow field lines. The location of the null point A corresponds to the one marked by the dashed circle in Fig. \ref{f4}(c). 
    Our simulation suggested that the X flares are produced here and this was also supported by several previous studies ({\it e.g.}, \citealt{2019A&A...628A.114P}, 
    \citealt{2020ApJ...890...10Z}, \citealt{2020ApJ...894...29B}). On the other hand, another null point B, which is further away form the null point A, is also reported 
    in  \cite{2018ApJ...869...69M}, \cite{2019ApJ...870...97Z}. This might have a potential to cause a flare and an eruption. In this section, we discuss a possibility of 
    flare triggering via reconnection in a different area than where the intruding motion was observed, such as  the null point B.
      
    To test this hypothesis, we perform another MHD simulation using the NLFFF reconstructed from the photospheric magnetic field, recorded at  02:36 UT on 
    September 6. This simulation is named "Run B", while the previous one is denoted "run A". In this study, the flare triggering reconnection in Run A is excited by 
    the numerical resistivity in the strong current region, due to the strong intruding motion of the negative polarity.  Figures \ref{f10}(a) and (b) exhibit the distribution 
    of $B_z$ observed at 02:36 UT and 08:36 UT, respectively, from which it is seen that the intrusion of the negative polarity has not fully taken place at 
    02:36 UT. Therefore, as of 02:36 UT, we expect that the reconnection should be suppressed at the region where the strong intruding motion was observed 
    later. Figures \ref{f10}(c) and (d) show the results of the magnetic twist, $T_w$, obtained from {the two} NLFFFs reconstructed at $t$=02:36 and 08:36, respectively 
    and  a quantitative comparison is shown in Fig.\ref{f10}(e). These distributions of $T_w$ are similar in both cases, but the value obtained at 08:36 UT is higher than 
    the one at 02:36 UT. Fig.\ref{f10}(f) shows the field lines at $t$=0.28 obtained from the MHD simulation when the NLFFF reconstructed at 02:36 UT is used as the 
    initial condition, in which we can find the two regions of steep gradient of the magnetic field marked by the dashed yellow and red  circles because $|\vec{J}|/|\vec{B}|$ 
    is strongly enhanced there. These locations are similar to those shown in Fig.\ref{f9} while the magnetic configuration at 02:36 UT is different from the one at 
    08:36 UT. From our results, the X2.2 flare appeared at the area indicated by the yellow circle when we used the photospheric magnetic field obtained 6 hours later.

% -------------------------------------------------------------------------------------------------------------------------------------------------------------------------------------------------------------------------------      
    \subsubsection{Two Different Solar Eruptions}
% --------------------------------------------------------------------------------------------------------------------------------------------------------------------------------------------------------------------------------
     The field lines constituting the MFR and  $|\vec{J}|/|\vec{B}|$ distribution on the vertical-cross section for Runs A and B are plotted in Figs.\ref{f11}(a) and (b), respectively. 
    In both the cases, the MFR is formed and the patterns of $|\vec{J}|/|\vec{B}|$ are similar to each other. In Run B, however, the location of the current sheet formed under 
    the MFR is slightly deviated from the region where the strong intruding motion is observed, while being enhanced above the region in Run A. Therefore, these results 
    suggest that the reconnection region is different in between Run A and Run B.  Specifically, the reconnection in Run B does not take place at the region intruded by the 
    negative polarity. Figures \ref{f11}(c) and (d) show snapshots of the 3D magnetic field lines obtained in Run B, from top and side views, respectively. The field lines 
    colored according to $|\vec{J}|/|\vec{B}|$ are traced from the region around the current sheet, following which the reconnection occurs outside of the highly twisted 
    lines. This location corresponds to the region indicated by red dashed circle in Fig.\ref{f10}(f) and the MFR can be created here while it takes longer time to make 
    the MFR compared to Run A.
      
    Interestingly, both Runs A  and B exhibited  eruptions albeit driven by different reconnections which take place at different locations: one is due to intrusion of the negative 
    polarity in Run A and the other is due to the reconnection at a different location, which is away from the region of the intruding motion of the sunspot, in Run B.  
    In this event, as shown in Figs. \ref{f4}(e) and (f), the strong brightening was observed at the beginning of the X2.2 flare at the region of the intruding motion of the sunspot. 
    Furthermore, from the Figs.\ref{f4}(a) and (b), the footpoints of the filed lines are anchored well on the strong brightening region of AIA 1600 \AA, which are not the field lines 
    connecting close to the null point of the magnetic field indicated by red dashed circle shown in Fig.\ref{f10}(f). If the eruption in Run B occurs, the flare brightening 
    observed in the AIA 1600 \AA \ observations would appear at the location outside of the highly twisted lines because they are surrounded by field lines participating 
    in the reconnection. 

    From the above, we confirm the possibility of another eruption which is different from the observed one, moreover, it would be possible 6 hours before the X2.2 
    flare was observed. \cite{2021ApJ...908..132Y} pointed out that the highly twisted field lines, which produce the X-flares, are already formed as of 2 days before and 
    suggested that the strong intrusion of the sunspot plays an important role for breaking the stable magnetic field through the reconnection. Since the value of electric 
    resistivity is considered as very small in the solar corona, the reconnection does not happen so easily in that situation unless the MFR becomes unstable or 
    the photospheric motion drives the upper magnetic field connecting the null point. In other words, if these would instead have taken place the area where the null 
    point B exists, a different eruption before the X-flares might have taken place.

%  -----------------------------------------------------------------------------------------------------------------------------------------------------------------------------------------------------------------
     \subsection{Gap Between the Observation and the Simulation}   
%  -----------------------------------------------------------------------------------------------------------------------------------------------------------------------------------------------------------------     
      We finally mention an inconsistency between the observations and the simulation. As shown in Fig.\ref{f1}(c),  the negative polarity starts the intrusion into the neighboring 
     positive polarity before the occurrence of the X2.2 flare.  From the detailed analysis of the observational data, \cite{2020ApJ...894...29B}  indicates that the intrusion 
     is a major candidate for driving the first flare (X2.2). When the negative polarity continuously intrudes into the neighboring opposite polarity, it can drive flux cancellation. 
     This creates the MFR through reconnection and eventually leads to the triggering of the second flare (X9.3). Although the overview is similar between the 
     observation and the simulation, we used a snapshot of the magnetic field in this study, which is taken before the X2.2 flare, as the boundary condition of our MHD 
     simulation. Thus we cannot discuss the phenomena along the temporal evolution of the photospheric magnetic field, {\it e.g.}, the energy buildup process after the X2.2 
     flare reported in \cite{2020ApJ...890...10Z}. In our simulation, the reconnection starting the X2.2 flare and after that is driven by numerical resistivity instead of the  
     photospheric magnetic cancellation through the intrusion.

% -------------------------------------------------------------------------  
% Two null points 
% -------------------------------------------------------------------------   
  \begin{figure}
  \epsscale{1.}
  \plotone{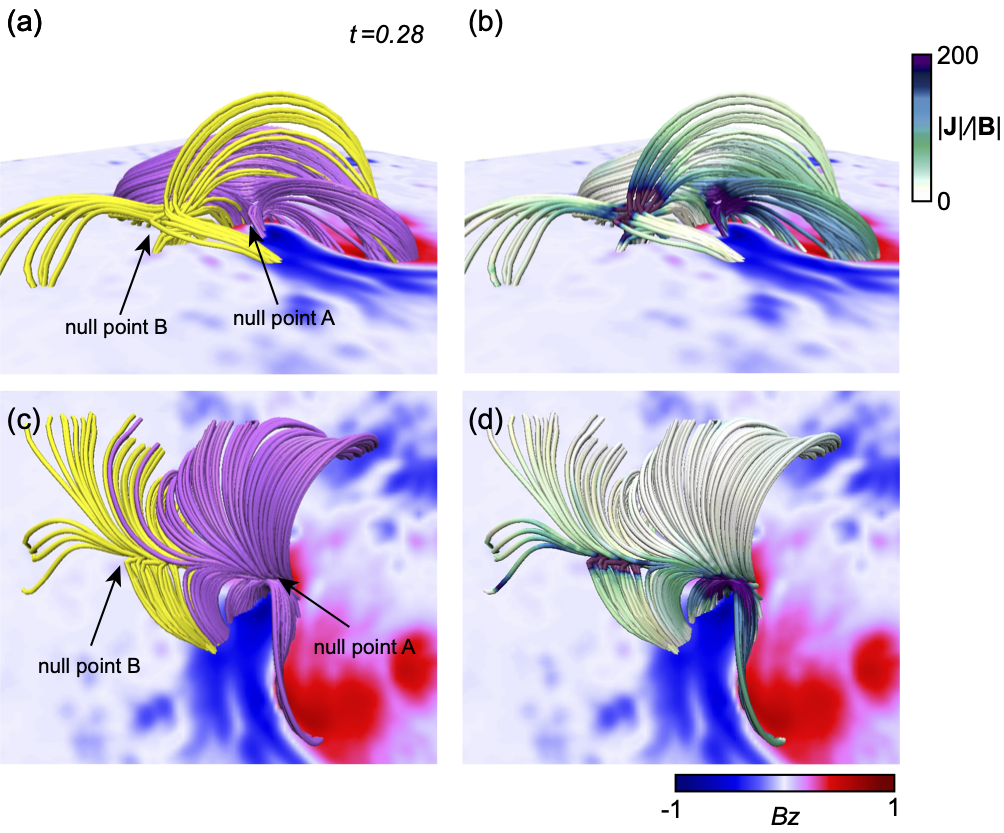}
  \caption{
                (a) The field lines at $t$=0.28 just before the X2.2 flare. The null point A is found in the purple field lines which correspond to those in Fig.\ref{f4}(c). Another null 
                      point B is shown by the yellow field lines. 
                (b) The magnetic field lines are colored by the value of $|\vec{J}|/|\vec{B}|$. The strong value is enhanced at each null point. 
                (c)-(d) The top view of (a) and (b).}
  \label{f9}
  \end{figure}
  \clearpage

% -------------------------------------------------------------------------  
% Compare between 08:36 UT and 02:36 UT
% -------------------------------------------------------------------------   
  \begin{figure}
  \epsscale{.9}
  \plotone{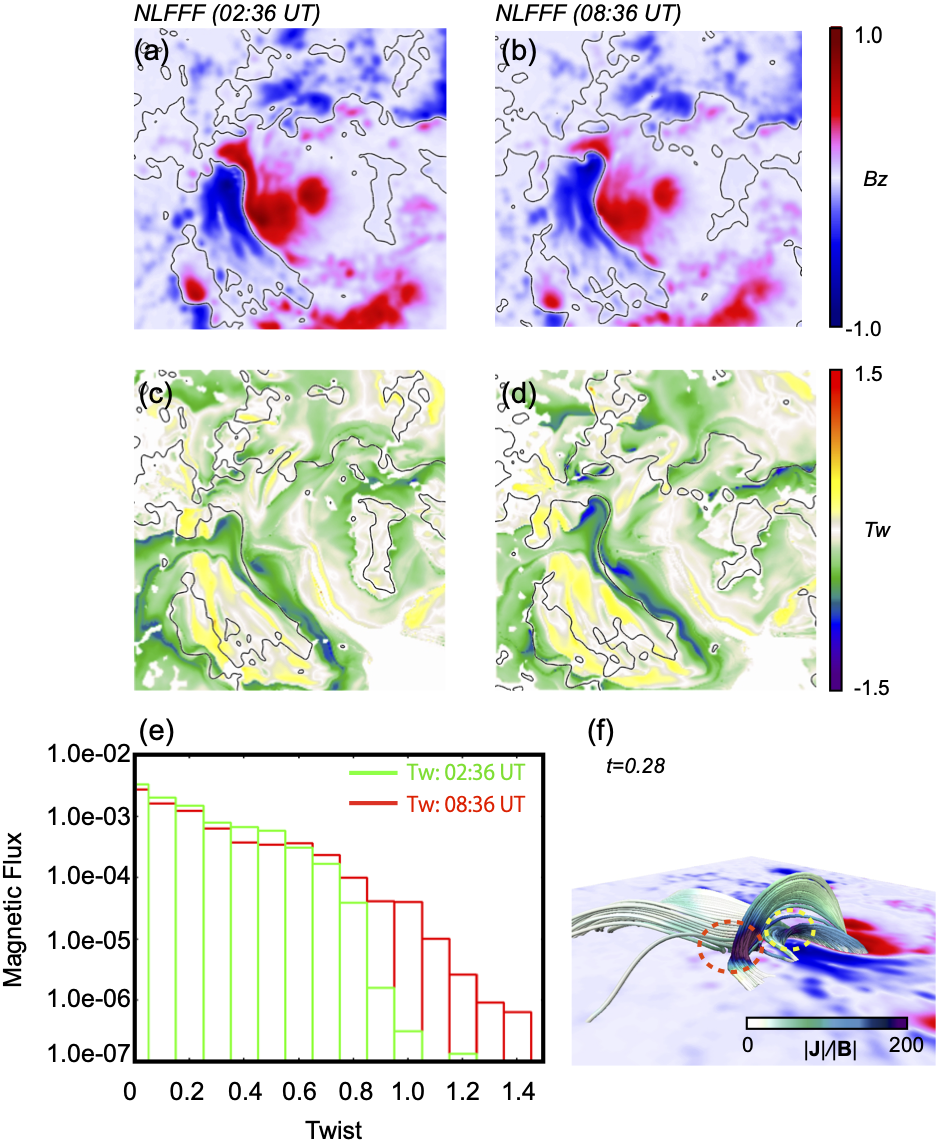}
  \caption{
                (a)-(b) $B_z$ distribution observed at $t$=02:36 UT and $t$=08:36 UT, respectively, where the black line corresponds to a contour of $B_z$=0.
                (c)-(d) Magnetic twist $T_w$ of each field line mapped on the photosphere at $t$= 02:36 UT and $t$=08:36 UT, respectively.
                      (e) Histogram of magnetic flux vs. twist where the green and red lines are calculated from the NLFFF reconstructed at $t$=02:36 UT and $t$=08:36 UT, respectively. 
                      (f)  The magnetic field obtained from the MHD simulation at $t$=0.28 when the NLFFF reconstructed at 02:36 UT is used. The color of the field lines correspond to 
                            the value of $|\vec{J}|/\vec{B}|$.}
  \label{f10}
  \end{figure}
  \clearpage
 
 % -------------------------------------------------------------------------  
% Dynamics of 3D Magnetic Flux Rope  
% -------------------------------------------------------------------------   
  \begin{figure}
  \epsscale{1.}
  \plotone{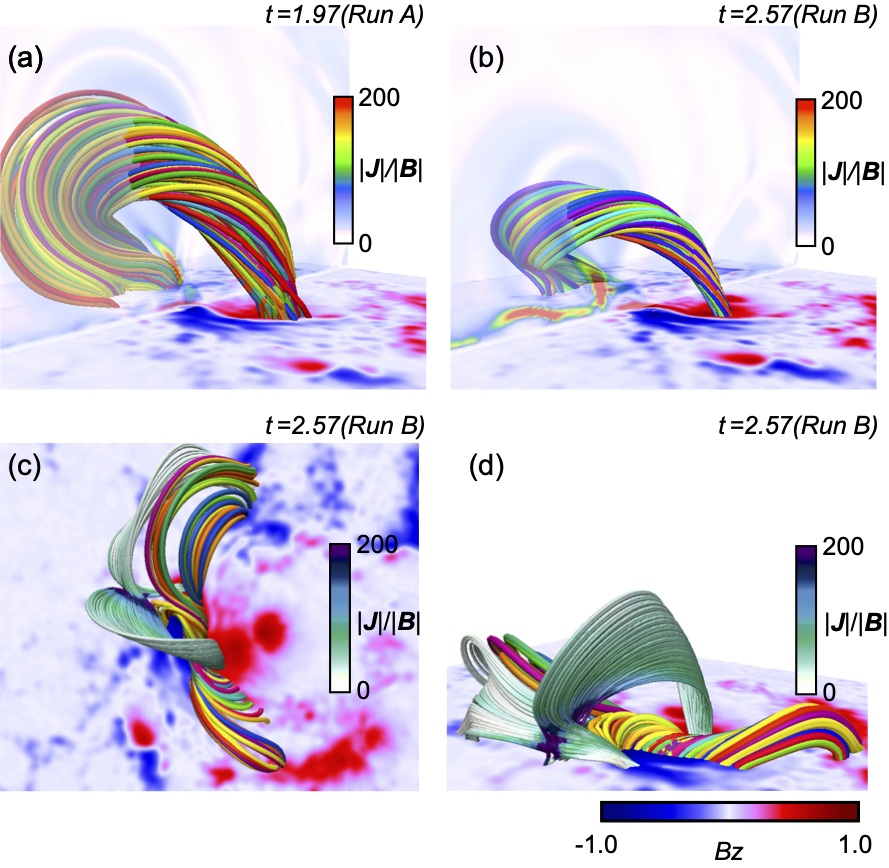}
  \caption{
                (a)-(b)  Erupting MFRs are plotted for Runs A and B together with $|\vec{J}|/|\vec{B}|$ along the vertical cross-section in each case. Run A corresponds to the 
                            simulation started with the NLFFF reconstructed at 08:36 UT, while Run B corresponds to that reconstructed at 02:36 UT. The location of the current 
                            sheet formed under the MFR is different between Run A and Run B. 
                (c)-(d)  The snapshots of magnetic field lines at $t$=2.57 in Run B from top and side views, respectively. The field lines drawn in the value of $|\vec{J}|/|\vec{B}|$, 
                            which are traced from the region around the current sheet, surround the highly twisted lines existing above PIL.}
  \label{f11}
  \end{figure}
  \clearpage

% ===============================================================================================================================
    \section{Summary}
% ===============================================================================================================================
    In order to reveal the MHD process of successive flares observed on September 6th 2017, we performed a data-constrained MHD simulation using the NLFFF, which is 
    reconstructed before the X2.2 flare, as the initial condition of the simulation. As shown in Fig.\ref{f5}, the twisted lines igniting the X2.2 flare are sandwiched by those 
    causing the X9.3 flare. The X2.2 flare removes a part of the innermost field lines, consequently, twisted lines existing at the outside form  the basis of the MFR through 
    reconnection as shown in the inset of Fig. \ref{f5}(b). The MFR is enhanced, in particular, its twist and size, and lifted up through a continuous reconnection after the X2.2 
    flare. The MFR eventually experiences the torus instability which triggers the eruption and gives rise to the X9.3 flare. In this simulation  we used one snapshot of the 
    photospheric magnetic field which is prior to the X2.2 flare, therefore we cannot trace the evolution of the photospheric motion taking place between the X2.2 to X9.3 
    flares and our simulation cannot take the phenomena derived from the time-dependent boundary motion, namely, the shearing and intruding motion of the negative 
    polarity. As further development, data-driven simulation(\citealt{2020ApJ...890..103T}) would be useful to cover the dynamics derived from the time-dependent boundary 
    condition.
    
     We confirmed another scenario for the eruption due to a reconnection, which takes place at away from the region where the intruding motion of the sunspot was observed,  
    by using the NLFFF 6 hours 30 minuets before the X2.2 flare.  Although this scenario hardly explains the AIA 1600 \AA \ observations, this result indicated that the eruption 
    might be achieved at the different than where the X2.2 flare was observed. \cite{2021ApJ...908..132Y} pointed out that the strong twist of the field lines are already 
    accumulated in the magnetic field 2 days before the X2.2 flare, so the intruding motion of the sunspot is important to break the stable magnetic field and cause the X2.2 flare.
    On the contrary, if some kind of disturbance is imposed on the region where the magnetic field connecting the null point exist, another eruption might happen.  As in this AR, 
    there are several possibilities for eruptions in ARs which have complex magnetic field. Therefore, it is important to detect the disturbance as the triggering process of the solar
    flare(\citealt{2013ApJ...778...48B}, \citealt{2017NatAs...1E..85W},  \citealt{2018ApJ...856...43B}, \citealt{2019ApJ...887..263K}) as well as understand the physical property of 
    the 3D magnetic structure(\citealt{2020Sci...369..587K}) and also magnetic topology for correctly understanding and predicting. In addition, as a further interesting issue, how 
    difference are large flares produced depending on a reconnection taken place at different area? We hope to address this topic in future work.

%\subsection{Animations}

%Authors may include animations in their articles.  A single still frame from 
%the animation should be included as a regular figure to serve as an example.
%The associated figure caption should indicate to the reader exactly what the
%animation shows and that the animation is available online.

%\begin{figure}
%\plotone{video3-eps-converted-to.pdf}
%\caption{Example image from the animation which is available in the electronic
%edition.}
%\end{figure}

\acknowledgments
 We are grateful to anonymous referee for many constructive comments and carefully checking the manuscript. 
 We thankl to Dr. Sung-Hong Park for supporting the CEA projection of the AIA images. SDO is a mission of NASA' s Living With a Star Program. This work was supported 
 by MEXT/JSPS KAKENHI, Grant Number JP15H05814 and MEXT as "Exploratory Challenge on Post-K computer" (Elucidation of the Birth of Exoplanets [Second Earth] and 
 the Environmental Variations of Planets in the Solar System). This work was also supported by "Nagoya University High Performance Computing Research Project for Joint 
 Computational Science" in Japan and the Center for Integrated Data Science of Institute of Space-Earth Environmental Research, Nagoya University.The research is partially 
 supported by the National Science Foundation under grant AGS-1954737. The visualization was performed by VAPOR(\citealt{2005SPIE.5669..284C}, 
 \citealt{2007NJPh....9..301C}).


\begin{thebibliography}{}
\bibitem[Amari et al.(1999)]{1999ApJ...518L..57A} Amari, T., Luciani, J.~F., Mikic, Z., et al.\ 1999, \apjl, 518, L57
\bibitem[Aulanier et al.(2012)]{2012A&A...543A.110A} Aulanier, G., Janvier, M., \& Schmieder, B.\ 2012, \aap, 543, A110
\bibitem[Bamba et al.(2013)]{2013ApJ...778...48B} Bamba, Y., Kusano, K., Yamamoto, T.~T., et al.\ 2013, \apj, 778, 48
\bibitem[Bamba \& Kusano(2018)]{2018ApJ...856...43B} Bamba, Y. \& Kusano, K.\ 2018, \apj, 856, 43
\bibitem[Bamba et al.(2020)]{2020ApJ...894...29B} Bamba, Y., Inoue, S., \& Imada, S.\ 2020, \apj, 894, 29
\bibitem[Bateman(1978)]{1978mit..book.....B} Bateman, G.\ 1978, Cambridge
\bibitem[Berger \& Prior(2006)]{2006JPhA...39.8321B} Berger, M.~A., \& Prior, C.\ 2006, Journal of Physics A Mathematical General, 39, 8321
\bibitem[Bobra et al.(2014)]{2014SoPh..289.3549B} Bobra, M.~G., Sun, X., Hoeksema, J.~T., et al.\ 2014, \solphys, 289, 3549
\bibitem[Clyne \& Rast(2005)]{2005SPIE.5669..284C} Clyne, J., \& Rast, M.\ 2005, \procspie, 284
\bibitem[Clyne et al.(2007)]{2007NJPh....9..301C} Clyne, J., Mininni, P., Norton, A., et al.\ 2007, New Journal of Physics, 9, 301
\bibitem[Dedner et al.(2002)]{2002JCoPh.175..645D} Dedner, A., Kemm, F., Kr{\"o}ner, D., et al.\ 2002, Journal of Computational Physics, 175, 645
\bibitem[Gordovskyy et al.(2014)]{2014A&A...561A..72G} Gordovskyy, M., Browning, P.~K., Kontar, E.~P., et al.\ 2014, \aap, 561, A72
\bibitem[Gordovskyy et al.(2020)]{2020ApJ...902..147G} Gordovskyy, M., Browning, P.~K., Inoue, S., et al.\ 2020, \apj, 902, 147
\bibitem[Guo et al.(2017)]{2017ScChD..60.1408G} Guo, Y., Cheng, X., \& Ding, M.\ 2017, Science China Earth Sciences, 60, 1408
\bibitem[Hou et al.(2018)]{2018A&A...619A.100H} Hou, Y.~J., Zhang, J., Li, T., et al.\ 2018, \aap, 619, A100
\bibitem[Inoue et al.(2014a)]{2014ApJ...780..101I} Inoue, S., Magara, T., Pandey, V.~S., et al.\ 2014, \apj, 780, 101
\bibitem[Inoue et al.(2014b)]{2014ApJ...788..182I} Inoue, S., Hayashi, K., Magara, T., et al.\ 2014, \apj, 788, 182
\bibitem[Inoue(2016)]{2016PEPS....3...19I} Inoue, S.\ 2016, Progress in Earth and Planetary Science, 3, 19
\bibitem[Inoue et al.(2018a)]{2018ApJ...867...83I} Inoue, S., Shiota, D., Bamba, Y., et al.\ 2018, \apj, 867, 83
\bibitem[Inoue et al.(2018b)]{2018NatCo...9..174I} Inoue, S., Kusano, K., B{\"u}chner, J., et al.\ 2018, Nature Communications, 9, 174
\bibitem[Janvier et al.(2013)]{2013A&A...555A..77J} Janvier, M., Aulanier, G., Pariat, E., et al.\ 2013, \aap, 555, A77
\bibitem[Jenkins et al.(2019)]{2019ApJ...873...49J} Jenkins, J.~M., Hopwood, M., D{\'e}moulin, P., et al.\ 2019, \apj, 873, 49
\bibitem[Jiang et al.(2018)]{2018ApJ...869...13J} Jiang, C., Zou, P., Feng, X., et al.\ 2018, \apj, 869, 13
\bibitem[Kang et al.(2019)]{2019ApJ...887..263K} Kang, J., Inoue, S., Kusano, K., et al.\ 2019, \apj, 887, 263
\bibitem[Kliem \& T{\"o}r{\"o}k(2006)]{2006PhRvL..96y5002K} Kliem, B., \& T{\"o}r{\"o}k, T.\ 2006, Physical Review Letters, 96, 255002
\bibitem[Kusano et al.(2020)]{2020Sci...369..587K} Kusano, K., Iju, T., Bamba, Y., et al.\ 2020, Science, 369, 587. doi:10.1126/science.aaz2511
\bibitem[Lemen et al.(2012)]{2012SoPh..275...17L} Lemen, J.~R., Title, A.~M., Akin, D.~J., et al.\ 2012, \solphys, 275, 17
\bibitem[Lin et al.(2020)]{2020ApJ...894...20L} Lin, P.~H., Kusano, K., Shiota, D., et al.\ 2020, \apj, 894, 20
\bibitem[Liu et al.(2018)]{2018ApJ...867L...5L} Liu, L., Cheng, X., Wang, Y., et al.\ 2018, \apjl, 867, L5
\bibitem[Liu et al.(2016)]{2016ApJ...818..148L} Liu, R., Kliem, B., Titov, V.~S., et al.\ 2016, \apj, 818, 148
\bibitem[Mitra et al.(2018)]{2018ApJ...869...69M} Mitra, P.~K., Joshi, B., Prasad, A., et al.\ 2018, \apj, 869, 69
\bibitem[Lumme et al.(2017)]{2017SoPh..292..191L} Lumme, E., Pomoell, J., \& Kilpua, E.~K.~J.\ 2017, \solphys, 292, 191. doi:10.1007/s11207-017-1214-0
\bibitem[Lysenko et al.(2019)]{2019ApJ...877..145L} Lysenko, A.~L., Anfinogentov, S.~A., Svinkin, D.~S., et al.\ 2019, \apj, 877, 145
\bibitem[Olmedo \& Zhang(2010)]{2010ApJ...718..433O} Olmedo, O., \& Zhang, J.\ 2010, \apj, 718, 433
\bibitem[Pesnell et al.(2012)]{2012SoPh..275....3P} Pesnell, W.~D., Thompson, B.~J., \& Chamberlin, P.~C.\ 2012, \solphys, 275, 3
\bibitem[Price et al.(2019)]{2019A&A...628A.114P} Price, D.~J., Pomoell, J., Lumme, E., et al.\ 2019, \aap, 628, A114
\bibitem[Priest \& Forbes(2002)]{2002A&ARv..10..313P} Priest, E.~R., \& Forbes, T.~G.\ 2002, \aapr, 10, 313
\bibitem[Romano et al.(2019)]{2019SoPh..294....4R} Romano, P., Elmhamdi, A., \& Kordi, A.~S.\ 2019, \solphys, 294, 4
\bibitem[Sakurai(1982)]{1982SoPh...76..301S} Sakurai, T.\ 1982, \solphys, 76, 301
\bibitem[Scherrer et al.(2012)]{2012SoPh..275..207S} Scherrer, P.~H., Schou, J., Bush, R.~I., et al.\ 2012, \solphys, 275, 207
\bibitem[Scolini et al.(2020)]{2020ApJS..247...21S} Scolini, C., Chan{\'e}, E., Temmer, M., et al.\ 2020, \apjs, 247, 21
\bibitem[Threlfall et al.(2018)]{2018SoPh..293...98T} Threlfall, J., Hood, A.~W., \& Priest, E.~R.\ 2018, \solphys, 293, 98
\bibitem[Toriumi, \& Wang(2019)]{2019LRSP...16....3T} Toriumi, S., \& Wang, H.\ 2019, Living Reviews in Solar Physics, 16, 3
\bibitem[Toriumi et al.(2020)]{2020ApJ...890..103T} Toriumi, S., Takasao, S., Cheung, M.~C.~M., et al.\ 2020, \apj, 890, 103
\bibitem[T{\"o}r{\"o}k et al.(2004)]{2004A&A...413L..27T} T{\"o}r{\"o}k, T., Kliem, B., \& Titov, V.~S.\ 2004, \aap, 413, L27
\bibitem[Vemareddy(2019)]{2019ApJ...872..182V} Vemareddy, P.\ 2019, \apj, 872, 182
\bibitem[Wang et al.(2018)]{2018ApJ...869...90W} Wang, R., Liu, Y.~D., Hoeksema, J.~T., et al.\ 2018, \apj, 869, 90
\bibitem[Wang et al.(2017)]{2017NatAs...1E..85W} Wang, H., Liu, C., Ahn, K., et al.\ 2017, Nature Astronomy, 1, 0085
\bibitem[Wiegelmann et al.(2006)]{2006SoPh..233..215W} Wiegelmann, T., Inhester, B., \& Sakurai, T.\ 2006, \solphys, 233, 215
\bibitem[Wiegelmann, \& Sakurai(2012)]{2012LRSP....9....5W} Wiegelmann, T., \& Sakurai, T.\ 2012, Living Reviews in Solar Physics, 9, 5
\bibitem[Wu et al.(2019)]{2019SoPh..294..110W} Wu, C.-C., Liou, K., Lepping, R.~P., et al.\ 2019, \solphys, 294, 110
\bibitem[Yamasaki et al.(2021)]{2021ApJ...908..132Y} Yamasaki, D., Inoue, S., Nagata, S., et al.\ 2021, \apj, 908, 132. 
\bibitem[Yan et al.(2018)]{2018ApJ...856...79Y} Yan, X.~L., Wang, J.~C., Pan, G.~M., et al.\ 2018, \apj, 856, 79
\bibitem[Yang et al.(2017)]{2017ApJ...849L..21Y} Yang, S., Zhang, J., Zhu, X., et al.\ 2017, \apjl, 849, L21
\bibitem[Zhong et al.(2019)]{2019ApJ...871..105Z} Zhong, Z., Guo, Y., Ding, M.~D., et al.\ 2019, \apj, 871, 105
\bibitem[Zou et al.(2019)]{2019ApJ...870...97Z} Zou, P., Jiang, C., Feng, X., et al.\ 2019, \apj, 870, 97
\bibitem[Zou et al.(2020)]{2020ApJ...890...10Z} Zou, P., Jiang, C., Wei, F., et al.\ 2020, \apj, 890, 10
\bibitem[Zuccarello et al.(2015)]{2015ApJ...814..126Z} Zuccarello, F.~P., Aulanier, G., \& Gilchrist, S.~A.\ 2015, \apj, 814, 126

\end{thebibliography}
\end{document}